\pdfoutput=1 
\documentclass[structabstract]{aa}  
\usepackage{amssymb}
\usepackage{textcomp}

\usepackage{float}
\usepackage{graphicx}
\usepackage{subfig}
\usepackage{txfonts}
\usepackage{natbib}
\usepackage{color}
\definecolor{rojo}{rgb}{1,0,0} 
\newcommand{\nc}{\newcommand}

\newcommand{\CII}{[C\,{\sc ii}]}

\newcommand{\OI}{[O\,{\sc i}]}

\newcommand{\HII}{H\,{\sc ii}}
\newcommand{\HI}{H\,{\sc i}}

\nc\micron{\mbox{$\mu$m}}
\nc{\cmcub}{\mbox{cm$^{-3}$}}
\nc{\cmsq}{\mbox{cm$^{-2}$}}
\nc{\Kkms}{\mbox{K~km~s$^{-1}$}}
\nc{\kms}{\mbox{km~s$^{-1}$}}
\nc{\mthirty}{\mbox{M\,33}}
\nc{\Tmb}{\mbox{$T_{\rm mb}$}}
\nc{\vlsr}{\mbox{v$_{\rm LSR}$}}
\nc{\twCO}{$^{12}$CO}
\nc{\thCO}{$^{13}$CO}
\nc{\msun}{\ensuremath{\mathrm{M}_\odot}}
\nc{\rsun}{\ensuremath{\mathrm{R}_\odot}}
\nc{\lsun}{\ensuremath{\mathrm{L}_\odot}}

\newcommand{\herschel}{{\it{Herschel}}}
\newcommand{\hermes}{{\tt HerM33es}}

\begin{document}
\title{Gas and dust cooling along the major axis of M\,33 ({\tt
    HerM33es})\thanks{\herschel\, is an ESA space observatory with
    science instruments provided by European-led Principal
    Investigator consortia and with important participation from
    NASA.}}  \subtitle{ISO/LWS \CII\ observations}

\author{ 
C.\,Kramer\inst{1}\and
J.\,Abreu-Vicente\inst{1}\and
S.\,Garc\'{i}a-Burillo\inst{2}\and
M.\,Rela\~{n}o\inst{3}\and
S.\,Aalto\inst{4}\and
M.\,Boquien\inst{5}\and			
J.\,Braine\inst{6}\and
C.\,Buchbender\inst{1}\and
P.\,Gratier\inst{6,7}\and
F.P.\,Israel\inst{8}\and
T.\,Nikola\inst{9}\and
M.\,R\"ollig\inst{10}\and
S.\,Verley\inst{3}\and
P.\,van der Werf\inst{8}\and
E.M.\,Xilouris\inst{11}
}

\institute{ 
     Instituto Radioastronom\'{i}a Milim\'{e}trica (IRAM), 
     Av. Divina Pastora 7, Nucleo Central, E-18012 Granada, Spain
\and 
     Observatorio Astron\'{o}mico Nacional (OAN) - Observatorio de Madrid, 
     Alfonso XII 3, 28014 Madrid, Spain
\and 
     Departamento de Fisica Te\'{o}rica y del Cosmos,
     Universidad de Granada, E-18071 Granada, Spain
\and
     Department of Radio and Space Science,
     Onsala Observatory, Chalmers University of Technology, 
     S-43992 Onsala, Sweden
\and 
     Aix Marseille Universit\'{e}, CNRS, LAM (Laboratoire d'Astrophysique 
     de Marseille) UMR 7326, 13388, Marseille, France
\and 
     Laboratoire d'Astrophysique de Bordeaux,  
     Observatoire de Bordeaux, Floirac F-33270, France
\and 
     IRAM, 300 rue de la Piscine, 38406 St. Martin d'H\`{e}res, France
\and 
     Leiden Observatory, Leiden University, PO Box 9513,  
     NL 2300 RA Leiden, The Netherlands
\and 
     Department of Astronomy, Cornell University, Ithaca, NY 14853, USA
\and
     KOSMA, I. Physikalisches Institut, Universit\"at zu K\"oln,   
     Z\"ulpicher Stra\ss{}e 77, D-50937 K\"oln, Germany    
\and 
     Institute of Astronomy and Astrophysics, National
     Observatory of Athens, P. Penteli, 15236 Athens, Greece
}
\date{}

 
  \abstract
  {} 
  {We aim to better understand the heating of the gas by observing the
    prominent gas cooling line \CII\ at 158\,$\mu$m in the
    low-metallicity environment of the Local Group spiral galaxy M\,33
    at scales of 280\,pc. In particular, we aim at describing the
    variation of the photoelectric heating efficiency with galactic
    environment.}
  { In this unbiased study, we used ISO/LWS \CII\ observations along
    the major axis of M\,33, in combination with Herschel PACS and
    SPIRE continuum maps, IRAM 30m CO 2--1 and VLA \HI\ data to study
    the variation of velocity integrated intensities. The ratio of
    \CII\ emission over the far-infrared continuum is used as a proxy
    for the heating efficiency, and models of photon-dominated regions
    are used to study the local physical densities, FUV radiation
    fields, and average column densities of the molecular clouds.}
  {The heating efficiency stays constant at 0.8\% in the inner
    4.5\,kpc radius of the galaxy where it starts to increase to reach
    values of $\sim3\%$ in the outskirts at about 6\,kpc radial
    distance. The rise of efficiency is explained in the framework of
    PDR models by lowered volume densities and FUV fields, for optical
    extinctions of only a few magnitudes at constant metallicity.  In
    view of the significant fraction of \HI\ emission stemming from
    PDRs, and for typical pressures found in the Galactic cold neutral
    medium (CNM) traced by \HI\ emission, the CNM contributes
    $\sim15\%$ to the observed \CII\ emission in the inner 2\,kpc
    radius of M\,33. The CNM contribution remains largely undetermined
    in the south, while positions between 2 and 7.3\,kpc radial
    distance in the north of M\,33 show a contribution of
    $\sim40\%\pm20\%$.  }
  {}


   \maketitle

%

\section{Introduction}

In photon dominated regions (PDRs), FUV photons from stars dominate
the chemistry and the energy balance in the interstellar gas.  All the
atomic and a large part of the molecular hydrogen of the ISM are
located in PDRs which emit an important fraction of far infrared (FIR)
and millimeter emission \citep{tielens-hollenbach1985,
  bakes-tielens1994, hollenbach-tielens1997}.

The \CII\ FIR fine structure line at 157.7\,$\mu$m is the most
important gas coolant. The \OI\ 63\,$\mu$m fine structure line starts
to dominate in denser and warmer regions when densities exceed about
$10^4$\,cm$^{-3}$ \citep{roellig2006}.  The photoelectric effect
provides one of the dominant gas heating processes in PDRs.  FUV
photons eject electrons from dust grains or PAH molecules, heating the
gas with their kinetic energy.  Theoretical models have predicted
efficiencies $\epsilon_{PE}$ of up to a few percent
\citep{weingartner-draine2001}, consistent with observations.  The
ratio of emerging \CII\ intensity over the infrared continuum radiated
by the dust has often been used as measure of this efficiency.
Observations of clouds in the Milky Way show variations over more than
2 orders of magnitude, between $10^{-4}$ and $3\,10^{-2}$
\citep[e.g.][]{vastel2001, habart2001, mizutani2004, jakob2007}. A
similar variation is found in observations of external galaxies
\citep[e.g.][]{malhotra2001, rubin2009}.  The scatter has been
attributed to changes in the mean charge of small grains and PAHs
\citep{okada2013}. However, the change of $\epsilon_{\rm {PE}}$ in
low-metallicity environments like those encountered in the Magellanic
Clouds or M\,33 is not yet well understood
\citep[e.g.][]{israel-maloney2011}. Interestingly, the efficiency
drops for local ULIRGs \citep{luhman2003,gracia-carpio2011} as well as
for some ULIRGs at high redshifts \citep[e.g.][]{stacey2010,cox2011}.
%
%

M\,33 is a nearby galaxy located at 840\,kpc distance
\citep{freedman1991}. Its overall metallicity is about half-solar
\citep{magrini2010}, only slightly larger than that of the Large
Magellanic Cloud (LMC) \citep{hunter2007}.  M\,33 is an Sc galaxy
exhibiting a prominent, flocculent spiral structure together with an
underlying extended diffuse component as seen e.g. in the 250\,$\mu$m
map of M\,33 conducted in the framework of the Herschel open time key
project \hermes\ \citep{kramer2010} (Fig.\,1).  M\,33 has a moderate
inclination of 56$^{\circ}$, allowing studies of the interstellar
medium with low depth along the line-of-sight.  Its proximity allows
high spatial resolution studies.

While previous studies have discussed \CII\ emission only at few
selected positions in M\,33, there has been no systematic study
describing the spatial variation of \CII\ emission in the disk of the
galaxy.  ~\citet{higdon2003} used ISO/LWS to study \CII\, and other
far-infrared (FIR) emission lines and the continuum in the nucleus and
six \HII\, regions.  They found a range of \CII/FIR$_{\rm LWS}$ values
of between 0.2\% and 0.7\%.  \citet{brauher2008} compiled ISO/LWS data
of 227 galaxies, including 23 positions in M\,33 with \CII\ data.
Plotting the \CII/FIR ratio for all galaxies they find variations
between $10^{-4}$ and somewhat more than 1\%.  \citet{mookerjea2011}
and \citet{braine2012} analyzed the first Herschel/PACS and HIFI
spectroscopic data sets of the \hermes\ project finding \CII/FIR
ratios between 0.01\% and 2\% 
%
in a $2'\times2'$ region centered on
the BCLMP\,302 \HII\ region and a ratio of 1.1\%
at the position of the BCLMP\,691 \HII\ region lying at
galacto-centric distances of 2.1 and 3.3\,kpc, respectively, along the
major axis.

In this paper, we present archival ISO/LWS \CII\ data along the major
axis of M\,33 out to 8\,kpc distance.  We study the radial
distributions and correlations between \CII\, and the FIR continuum,
CO, \HI, and H$\alpha$. We also study the radial distribution of the
\CII/FIR ratio. We compare the observations in M33 with data of star
forming regions in the Milky Way, with other external galaxies,
including low-metallicity objects.  Local volume densities and FUV
fields of the \CII\ and CO emitting gas are estimated using the
\citet[][K99]{kaufman1999} PDR model. The observed \CII\ emission is
also compared with an estimate of the \CII\ emission emitted by atomic
clouds.

\begin{figure}
\includegraphics[scale=0.5]{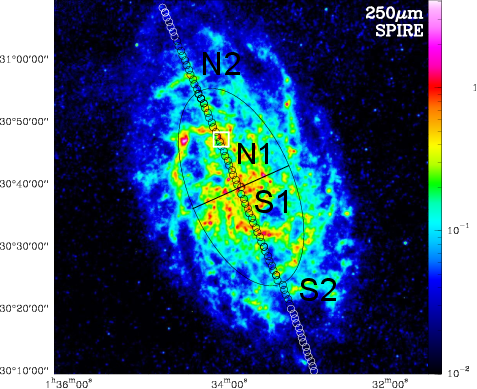}
\caption{Herschel SPIRE 250\,$\mu$m map of M\,33
  \citep{xilouris2012}.  Units are Jy/($18''$\,beam).  Circles mark
  the position and beam size of the ISO/LWS \CII\ observations along
  the major axis of M\,33.  The ellipse delineates the galacto-centric
  distance of 4.5\,kpc, dividing the observed inner (N1 and S1) and
  outer (N2 and S2) regions. The white square marks the BCLMP\,302
  \HII\ region.  Coordinates are R.A. and Dec. (Eq.\,J2000).}
\label{Fig1}
\end{figure}

\begin{table*}
\caption{Observation Log of all \CII\ observations of ISO/LWS along the major axis of M\,33}             
\centering                          
\begin{tabular}{c c c c c c c c c}        
\hline\hline                 
IDA Name & Abbrev.\tablefootmark{a} & TDT\tablefootmark{b} (Obs. {\verb # }ID) & R.A.\tablefootmark{c} & Dec.\tablefootmark{c}  & Observer ID & Ref. 
& Type & AOT\\    
\hline                        
 M\,33S2 & S2 & 59901107 & 01h33m08.5s & +30d17m00.0s & KMOCHIZU & $-$          & Raster & LO2\\
 M\,33S  & S1 & 78600403 & 01h33m37.1s & +30d31m34.7s & KMOCHIZU & (1) 
& Raster & LO2\\
 M\,33 Nucleus & & 80800367 & 01h33m50.9s & +30d39m36.8s & HSMITH & (1, 2) & Point & LO2\\
 M\,33N  & N1 & 78600801 & 01h34m07.3s & +30d46m55.7s & KMOCHIZU & $-$ 
& Raster & LO2\\
 M\,33N2 & N2 & 59900605 & 01h34m36.3s & +31d01m29.2s & KMOCHIZU & $-$        
& Raster & LO2\\
\hline
\label{Tab1} 
\end{tabular}\\
\tablefoottext{a}{Abbreviation used in the text.} 
\tablefoottext{b}{Target Dedicated Time.}
\tablefoottext{c}{The given coordinates are the central
  position of each raster strip of 19 positions.}
\tablebib{
  (1) \citet{brauher2008},
  (2) \citet{higdon2003}.
}
\end{table*}

\section{Observations and data analysis}\label{observations}

\subsection{\CII\ $158\mu$m (ISO/LWS)}\label{CII-obs}

We list all ISO\footnote{Infrared Space Observatory
  \citep{kessler1996}}/LWS\footnote{Long Wavelength Spectrometer in
  ISO \citep{gry2003}} \CII\ spectra observed along the major axis of
M\,33 at a position angle of $23^\circ$ in Table\,\ref{Tab1}. The
spectra were observed using the partial grating scan mode (LWS
AOT\footnote{Astronomical Observing Template} LO2). This AOT covers
the wavelength range $43-196.9$ $\mu$m and it has a medium spectral
resolution of $\Delta\lambda/\lambda\sim$ 200 corresponding to
1500\,\kms\ at the wavelength of the \CII\ line.  The LWS flux
calibration and its relative spectral response function were derived
from observations of Uranus ~\citep{swinyard1998}. The angular
resolution is $69.4''$ ~\citep{gry2003} which corresponds to a linear
resolution of 280\,pc. The spectra had been automatically processed by
the ISO system, passing scientific validation. We retrieved spectra at
77 positions from the ISO Data Archive (IDA) for further processing.
The observed positions cover a range of about $\pm\,8\,$kpc ($\pm33'$)
from the nucleus on a grid of about 208\,pc (Figure~\ref{Fig1}).

We averaged the observations at each position, subtracted linear
baselines, and fitted a Gaussian to the line. Data were analyzed using
the ISO Spectral Analysis Package \citep[ISAP v2.1][]{sturm1998}. In
the following, we assume a calibration error of
15\%~\citep{higdon2003}. 
%
Some sample spectra are shown in Appendix\,\ref{sec-examples}.

Figure\,\ref{distribution} shows the variation of \CII\ intensities
along the major axis. \CII\ is detected above 3$\sigma$ at 36
positions. Stacking of neighboring positions increased the number of
detections. Horizontal errorbars indicate the region over which the
data were stacked. Tables\,\ref{PosSur}, \ref{PosNor} in the Appendix
list the stacked positions.

\subsection{Far-infrared continuum} \label{FIR-obs} 
   
To measure the total FIR continuum, we combined SPIRE and PACS maps of
M\,33 at five wavelengths between 500 and 100$\,\mu$m wavelength
\citep{boquien2011, xilouris2012}, taken in the framework of \hermes,
with MIPS/Spitzer 24 and 70\,$\mu$m maps \citep{verley2007,
  tabatabaei2007}.

These maps were smoothed to the ISO/LWS resolution using Gaussian
kernels.  The fluxes were extracted using circular apertures of
69.4$''$ centered at the ISO/LWS positions.  A two-component grey body
function was fitted to the spectral energy distribution (SED) at each
position following the method described in \citet{kramer2010}.
Integrating between 42.5 $\mu$m and 122.5$\mu$m \citep{dale-helou2002}
yields the FIR surface brightness.  The dust emissivity index $\beta$
was fixed at 1.5, which was found to be the best-fitting value for
M\,33 \citep{kramer2010, xilouris2012}.  The total FIR luminosity is,
however, robust against changes of $\beta$.
Figure\,\ref{distribution} shows the variation of relative FIR
intensities along the major axis (cf.\,Table~\ref{Intensities}). A few
sample SEDs are shown in Appendix\,\ref{sec-examples}.
 
We also integrated the fitted SEDs over the range 3-1000\,$\mu$m
wavelengths to estimate total infrared (TIR)
intensities~\citep{dale-helou2002}, thereby deriving the ratios
between TIR over FIR which lie between 2.7 and 1.3
(Table~\ref{Intensities}). 

\subsection{CO, \HI, H$\alpha$} \label{comp-obs}

Complementary CO and \HI\ data were used as
tracers of the molecular and atomic gas to compare with \CII\
emission. 

The CO 2--1 line was mapped with the IRAM 30m telescope by
\citet{gardan2007} and \citet{gratier2010}. These maps cover the major
axis out to a distance of 8.5\,kpc in the north and 6.5\,kpc in the
south. The CO map covers all ISO/LWS \CII\ positions but the eight
southern-most positions.

We determined $3\,\sigma$ upper limits of the integrated intensities
using $\sigma = \sqrt{N}\Delta v_{\rm res}\,T^{\rm rms}_{\rm mb}$ with
the number of channels $N$ over the velocity extent of the line, the
velocity resolution $\Delta v_{\rm res}$ and corresponding baseline
rms $T^{\rm rms}_{\rm mb}$.

The \HI\, VLA map of M\,33 \citep{gratier2010} covers the entire
galaxy out to 8.5 kpc radial distance. \HI\ is detected at all ISO/LWS
positions. While no single dish data were combined with the
interferometric observations, the total flux recovered over the entire
galaxy by the interferometric observations alone corresponds to more
than 90\% of the flux measured at the Arecibo single dish telescope
\citep{putman2009}.

We also used a map of H$\alpha$ emission presented
in~\citet{hoopes-walterbos2000} and by ~\citet{verley2007}.  These
data were obtained at the 0.6\,m Burrell-Schmidt telescope at Kitt
Peak National Observatory (KPNO).  H$\alpha$ is detected at 42 ISO/LWS
positions. We have calculated upper limits by measuring rms levels of
the map in regions without H$\alpha$ emission close to ISO/LWS \CII\
positions. 

Intensities have been calculated by smoothing all data to the angular
resolution of the LWS \CII\ data as described in the
Appendix\,\ref{co10}.

The variation of relative intensities of CO, \HI, and H$\alpha$ is
shown in Figure\,\ref{distribution} and absolute intensities are
listed in Table~\ref{Intensities}. We stacked CO, \HI, and H$\alpha$
over the same positions as \CII\ (cf.\,Tables\,\ref{PosSur},
\ref{PosNor}).

\section{Results}

\subsection{Correlation between \CII\ and FIR, H$\alpha$, CO, \HI} 
\label{radial}

\begin{figure}
\centering
\includegraphics[angle=0,scale=0.5]{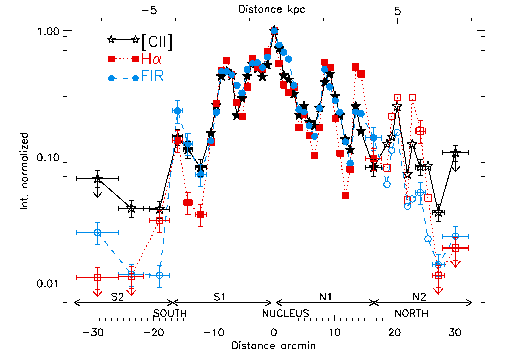}
\includegraphics[angle=0,scale=0.5]{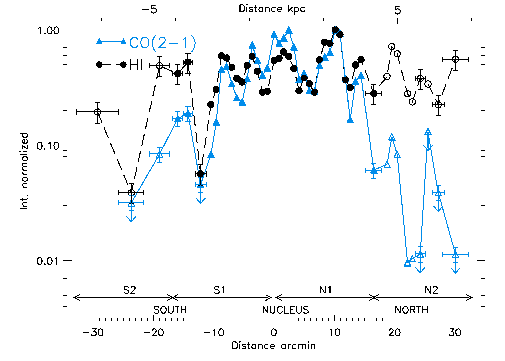}
\caption{\CII, FIR, H$\alpha$, CO 2--1, and \HI\
  normalized integrated intensities along the major axis of M\,33.
  Closed symbols show data within 4.5\,kpc of the nucleus (N1, S1) and
  open symbols show observations in the outer galaxy (N2, S2).
  Horizontal error-bars show the region over which different \CII\
  spectra from neighboring positions were averaged.  }
\label{distribution}
\end{figure}

\begin{figure}
\centering
\includegraphics[angle=0,scale=0.5]{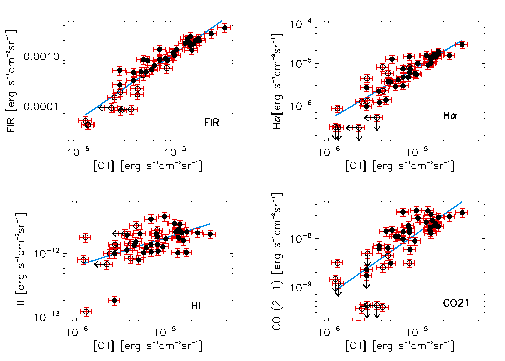}
\caption{Correlation of \CII\ intensities with those of the FIR
  continuum, H$\alpha$ emission, CO, and \HI\ in double-logarithmic
  plots. All intensities are given in units of
  erg\,s$^{-1}$\,cm$^{-2}$\,sr$^{-1}$. Straight lines delineate the
  results of linear least squares fits to the data. Closed and open
  symbols distinguish between positions in the inner (S1, N1) and
  outer (S2, N2) disk of M\,33, respectively. }
\label{correlaciones}
\end{figure}

The emission of \CII, the FIR continuum, and H$\alpha$, are all well
correlated along the major axis, especially in the inner $10'$
(Fig.~\ref{distribution}). These tracers of star formation all peak at
the nucleus and drop by more than one order of magnitude beyond
$\sim20'$ radial distance. Closer inspection shows that the FIR
continuum drops steeper than \CII, a finding which is discussed
further below. In general however, \CII, FIR, and H$\alpha$ all trace
the spiral arms as well as the inter-arm regions. The close
correlation is even more clearly seen in Figure\,\ref{correlaciones}
where we plot the various tracers against \CII\ emission. A close
correlation of \CII\ with other tracers of star formation is well
known for other sources, and has been seen e.g. also in the ISO/LWS
maps of a portion of the northern arm of M\,31 by
\citet[][]{rodriguez-fernandez2006}.

The atomic and molecular gas traced by \HI\ and CO also drops with
radial distance, however the drop is much steeper for CO than for
\HI\, (Fig.~\ref{distribution}).  In the inner part of the galaxy
within $10'$ radial distance, the distribution of CO and \HI\ is
rather flat, in contrast to that of e.g. the FIR continuum. The
distribution of emission of both gas tracers is not symmetrical with
respect to the nucleus. Instead, CO and \HI\ peak near $10'$ ($\sim
2.5$ kpc) to the north and only show a secondary maximum at the
nucleus.  The absolute maximum corresponds to GMC\,91
\citep{engargiola2003,gratier2010,buchbender2013} and to cloud 245 in
\citet{gratier2012}.

\begin{table}[h]
  \caption{Linear least squares fits to the correlations between 
 \CII\ and FIR, H$\alpha$, CO, \HI\ (cf.\,Fig.~\ref{correlaciones}) of the
form $\log$FIR$=a+b\,\log$\CII\ with correlation coefficients $r$.}
%
\centering 
\begin{tabular}{ccccc}
\hline\hline          
     & FIR-\CII & H$\alpha$-\CII & CO(2-1)-\CII & \HI-\CII  \\ \hline 
 $a$ & 3.71     & 1.08             & $-1.52$    & $-9.74$       \\
 $b$ & 1.33     & 1.23             & 1.27       & 0.41                   \\
 $r$ & 0.90     & 0.79             & 0.59       & 0.30                   \\
\hline
\label{tabcor}     
\end{tabular}
\end{table} 
%
%

Figure ~\ref{correlaciones} shows correlations between \CII\ emission
and FIR, H$\alpha$, CO, and \HI.  Linear fits are weighted by the
errors along both axes which we assume to be 15\% for \CII\ and \HI,
and 20\% for CO and the FIR continuum. The fits confirm that \CII\ is
strongly correlated with FIR and H$\alpha$, with linear correlation
coefficients ($r$) 0.90 and 0.79, respectively (Table~\ref{tabcor}).
In contrast, the correlation with CO is much weaker, $r$=0.59, and
very poor for \HI, $r$=0.30.  This confirms that \CII\ is a good
tracer of the star formation rate at the ISO/LWS beam size scale.
Points from the northern most region N2 deviate slightly from the fit
both in the \CII-FIR and the \CII-H$\alpha$ plots, showing a worse
correlation between \CII\ and star forming tracers in the northern,
outer galaxy.

\begin{figure}
\centering
\includegraphics[angle=0,scale=0.5]{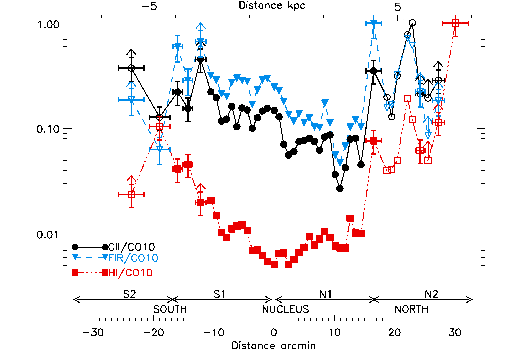}
\caption{Normalized radial distributions of the
  \CII/CO 1--0, FIR/CO 1--0, and \HI/CO 1--0 intensity ratios on the
  erg-scale. }
\label{ciico-dist}
\end{figure}

\begin{figure}
\centering
\includegraphics[angle=0,scale=0.5]{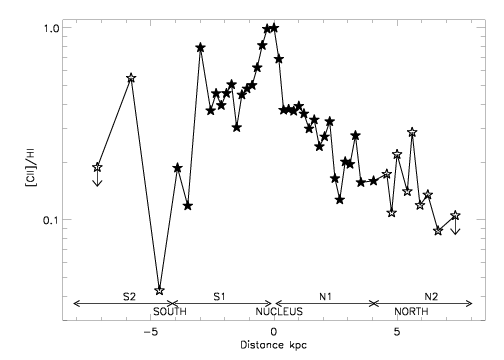}
\caption{ Normalized radial distribution of the \CII/\HI\ intensity
  ratio. }
\label{ciihi-dist}
\end{figure}

\subsection{Radial variation of intensity ratios}
      
The radial distributions of the ratios \CII/CO 1--0, FIR/CO 1--0 and
\HI/CO 1--0 are shown in Figure~\ref{ciico-dist}. For this, we
estimated CO 1--0 intensities from the CO 2--1 line
(cf.\,Appendix~\ref{co10}).  All three ratios show a minimum in the
inner parts of the galaxy and an increase towards the outer parts.
This reflects the steep drop of CO intensities seen already in the
radial distribution of intensities (Fig.\,\ref{distribution}). The
\HI/CO ratio shows a minimum near the nucleus and rises steadily
towards the outskirts over about two orders of magnitude. While this
behavior is about symmetrical in the North and South of the galaxy,
this symmetry is broken for the \CII/CO and FIR/CO ratios. The latter
two ratios show a minimum at about $10'$ to the North near GMC\,91
where the CO emission peaks (Fig.\,\ref{distribution}). To the south
of this minimum, the two ratios steadily increase. Towards the north,
the two ratios also increase, but with larger scatter.

In general, the \CII\ emission drops more steeply with galacto-centric
radius than the \HI\ emission. This is most clearly seen in the
northern part of the strip (Fig.\,\ref{ciihi-dist}). In contrast, the
southern outer disk shows strongly varying \CII/\HI\ ratios owing to
the strong variability of \HI\ emission
(cf.\,Fig.\,\ref{distribution}). 

\begin{figure}
\centering
\includegraphics[angle=0,scale=0.5]{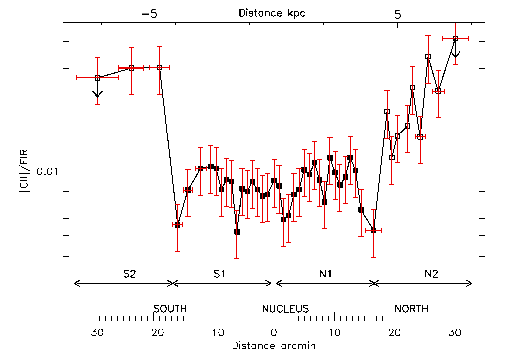}
\caption{Radial distribution of the \CII/FIR ratio along the major axis
  of M\,33. }
\label{pehe-dist}
\end{figure}

\begin{figure}
   \centering
   \includegraphics[angle=0,scale=0.5]{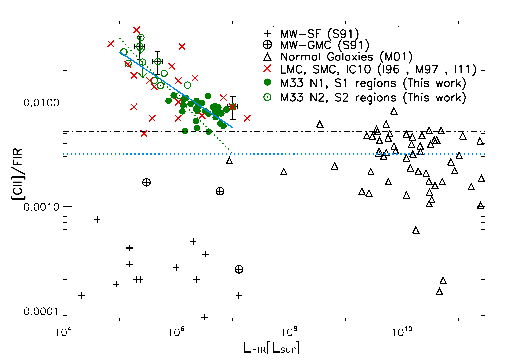}
   \caption{\CII/FIR ratio as function of $L_{\rm {FIR}}$. The blue
     solid line shows a linear fit to the whole data set of M\,33
     while the green dotted line shows a linear fit to N2 and S2
     points of the outer galaxy only.  The black dashed-dotted
     horizontal line shows the lower \CII/FIR value found in our data
     set of M\,33. The blue dotted horizontal line shows the average
     \CII/FIR value in normal galaxies~\citep[][M01]{malhotra2001}. We
     also show data of Milky Way regions \citep[][S91]{stacey1991} and
     the low metallicity objects LMC, SMC, IC\,10
     \citep[][I96]{israel1996}, ~\citep[][I11]{israel-maloney2011},
     ~\citep[][M97]{madden1997}.}
        \label{pehe}
\end{figure}

\subsection{\CII /FIR ratio}
\label{CII-FIR-correlation}

In the inner part of the galaxy (N1, S1), the \CII /FIR ratio stays
constant within the measuring accuracy at about 0.8\% $\pm$ 0.2\%
(Fig.\,\ref{pehe-dist}), but rises significantly and steeply in the
outskirts to values of $\sim3\%$.  The rise of the \CII/FIR ratio is
caused by the steep drop of FIR emission relative to the \CII\
(Fig.~\ref{distribution}), as is also seen in a plot of \CII/FIR
versus FIR (Fig.~\ref{pehe}).  The abrupt increase of the \CII/FIR
ratio at about 4.5\,kpc radial distance, both in the north and in the
south of the major axis, occurs where the morphology of the whole
galaxy seen in the optical changes \citep[e.g.][]{sharma2011}, i.e.
just beyond the location of the prominent spiral arms.
It also occurs in a region where the \HI/CO ratio rapidly increases
(Fig.\,\ref{ciico-dist}).  As the O/H abundance gradient in M\,33 is
shallow with a slope of only $\sim-0.035$\,dex\,kpc$^{-1}$ and with no
signs of a break \citep{magrini2010}, the metallicity cannot be key
for the sudden increase of the \CII/FIR ratio.

Though only few studies of the radial variation of the \CII/FIR ratio
in galaxies exist, observations of M\,31, M\,51, NGC\,6946, and the
Milky Way indicate that in general this ratio rises with
galacto-centric distance. See the discussion in
\citet{rodriguez-fernandez2006} who present ISO/LWS maps of a portion
of the northern arm of M\,31 at 12\,kpc radial distance showing a
rather constant and high ratio of 2\% while the nucleus shows only
0.6\%. As M\,31 is at the same distance as M\,33, the LWS observations
of M\,31 sample the same linear scale as the present observations of
M\,33.  Further below, we attempt to interpret the radial variation
found along the major axis of M\,33 using PDR models.

In M33, the \CII/FIR ratio drops with increasing FIR luminosity.  A
power law fit to the entire data set results in \CII/FIR $\propto$
$L_{\rm{FIR}}^{-0.34}$ ($r$=0.83). A fit to the data of the outer disk
only results in a slightly steeper slope (Fig.~\ref{pehe}).  In this
figure, we also compare the M\,33 data with those of other galaxies.
The M\,33 data are located in the region of low FIR luminosities
$L_{\rm {FIR}}$=10$^5$--10$^7$\,L$_{\sun}$ relative to those of the
normal galaxies observed by \citet{malhotra2001} and high \CII/FIR
ratios of between 0.6\% and 3\%.  These data lie in the same region as
the data of the LMC, SMC and IC10.

The far-infrared luminosities of the low-metallicity systems are not
directly comparable as their distances differ. However, while the LMC
and SMC data were taken at 14 to 16\,pc linear resolutions,
respectively \citep{israel1996, israel-maloney2011}, the \CII\ and FIR
values were taken from maps where the fluxes were averaged over
individual sources of between 35 and 70\,pc in size. The IC\,10
observations have a resolution of 290\,pc \citep{madden1997}, similar
to that of the M\,33 ISO/LWS data.

The M\,33 data in Figure~\ref{pehe} lie in the high limit of \CII/FIR
ratios found in a sample of normal galaxies studied with ISO/LWS by
\citet{malhotra2001}. The latter appear on the right side of the plot
with \CII/FIR = 0.01\% - 0.7\% and $L_{\rm{FIR}} = 10^7-10^{11}$\,L$_{\sun}$
showing an average \CII/FIR ratio of 0.32\%.

The \CII/FIR ratios in M\,33 are 25 -- 200 times higher than in Milky
Way star forming regions for a similar $L_{\rm{FIR}}$ range, while
this factor decreases an order of magnitude (2.5 -- 20) in Galactic
GMCs.

\begin{figure}[h]
   \centering
        \includegraphics[scale=0.5,angle=0]{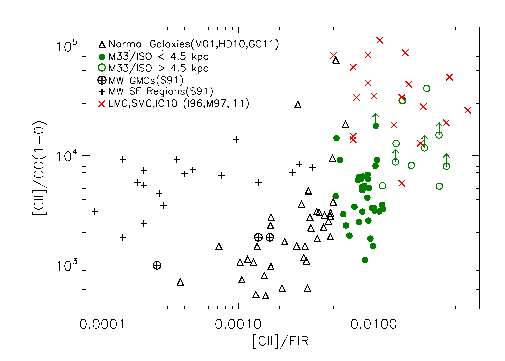}\\
        \includegraphics[scale=0.5,angle=0]{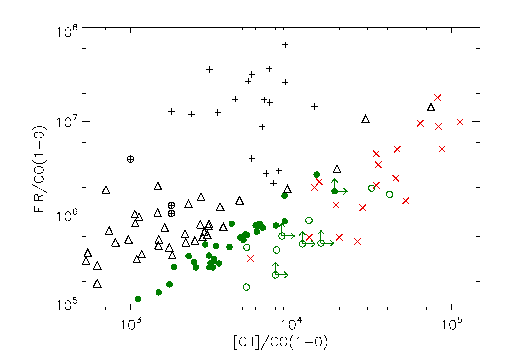}\\
        \caption{Diagnostic plots of ratios of \CII, CO and the FIR
          continuum for M\,33 and four other types of objects,
          low-metallicity systems, normal galaxies, Milky Way star
          forming regions, and GMCs. {\bf Upper panel:} \CII/CO 1--0
          vs. \CII/FIR. {\bf Bottom panel:} FIR/CO 1--0 vs. \CII/CO
          1--0. All panels are shown on a logarithmic scale.  Symbols
          used have the same meaning than those used in
          Figure~\ref{pehe}. }
   \label{israel}             
\end{figure}

\subsection{M\,33: A bridge between dwarf and normal galaxies ?}

After comparing \CII/FIR differences between M\,33 and normal galaxies
we also compared CO/FIR and \CII/CO ratios found in M\,33 with those
of other sources (Fig.~\ref{israel}).

In these plots, we find four mostly disjunct groups of ratios. There
is a group formed by low metallicity objects and M\,33 of the outer
galaxy (N2, S2).  The M\,33 points of the inner galaxy (N1, S1) form
another group, which seems to connect low metallicity objects with
normal galaxies, and Milky Way GMCs (S91) which form the third group.
Finally, PDRs in the Milky Way exposed to high FUV fields (MW SF
regions, S91) form the fourth group.  For normal galaxies, the \CII/CO
1--0 ratio increases with \CII/FIR.  For M\,33 and other low
metallicity objects the FIR/CO 1--0 ratios increase strongly with
increasing \CII/CO 1--0, a correlation which is also seen for normal
galaxies though with increased scatter.  The data points of the inner
disk of M\,33 lie in-between the data of the normal galaxies at low
\CII/CO ratios and the data of the other low-metallicity systems
showing high \CII/CO ratios.

On average, the inner parts of M\,33 (N1, S1) show lower \CII/CO and
FIR/CO ratios than the other low-metallicity objects.  This suggests
that CO is less photo dissociated in the inner disk of M33 than in the
outer regions (S2, N2) and in the other low-metallicity objects.  This
is further discussed in the next paragraphs when comparing the
observed ratios with PDR models.
%

\begin{figure*}
\centering
\includegraphics[angle=0,scale=1.]{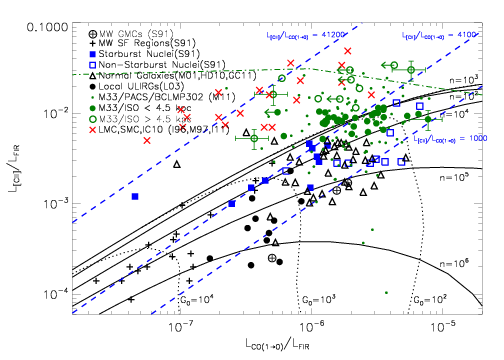}
\caption{ \CII\ versus CO, normalized with the FIR continuum. Big
  green filled circles show ISO/LWS data of the inner S1, N1 regions
  of M\,33 while open circles show data of the outer S2, N2 regions.
  Small green circles show PACS observations of the BCLMP\,302 \HII\,
  region in M\,33 (M11). In addition we show data from Milky Way GMCs
  and star forming regions, and from other galaxies, compiled from
  ~\citet[][S91]{stacey1991}, ~\citet[][HD10]{hailey-dunsheath2010},
  ~\citet[][GC11]{gracia-carpio2011}(GC11),
  ~\citet[][I96]{israel1996}, ~\citet[][I11]{israel-maloney2011},
  ~\citet[][M97]{madden1997}, ~\citet[][M01]{malhotra2001}, and
  ~\citet[][L03]{luhman2003}. The lowest \CII/CO ratio observed with
  ISO/LWS in M\,33 is 1000 (lower blue dashed line), while the highest
  ratio is 41200 (upper blue dashed line).  Black solid and dotted
  lines indicate lines of constant density $n$ and FUV field $G_0$,
  respectively, from the standard K99 PDR model with $A_{\rm
    {V}}=10$\,mag and solar metallicity $Z=1$. The dashed-dotted green
  line shows a K99 PDR model result for $A_{\rm {V}}=1$\,mag, $Z=1$,
  $n=10^3$\,cm$^{-3}$.  The knee of this curve is for $G_0=10^{0.5}$
  (cf.\,Fig.\,18 in K99).}
%
%
\label{hailey-todo}
\end{figure*}

\section{Discussion}

\begin{table} 
  \caption{Range of densities and FUV fields consistent with the
standard K99 PDR model for sources shown in the diagnostic plot of
\CII/FIR vs. CO/FIR (Fig.~\ref{hailey-todo}).}
\centering 

\begin{tabular}{cccc}
  Objects 	       & \CII/(CO 1--0) 
                                       & $n$ [cm$^{-3}$]      & $G_0$		\\\hline\hline
  M\,33 (N1-S1)        & 1000 -- 8000  & $<10^4$ & $<10^3$		\\
  M\,33 (N2-S2)        & 5000 -- $4.1\,10^4$ & --- 		      & ---			\\
  MW-SF 	       & 1000 -- 5000  & 10$^2$--$10^6$       & 10$^3$ -- $>$ 10$^4$    \\
  MW-GMCs 	       & 500 -- 1000   & 10$^5$--$10^6$       & 500 -- $\gtrsim 10^3$   \\
  Normal galaxies      & 500 -- 4000   & 10$^3$--$10^6$       & 10$^2$ -- 10$^3$ 	\\
  ULIRGs	       & 500 -- 4000   & 10$^4$--$10^6$       & 10$^3$ -- 10$^4$	\\
  Starburst nuclei     & $\sim 4100$   & 10$^3$--$10^4$       & 500 -- $\gtrsim 10^3$   \\
  Non-starburst nuclei & 500 -- 4000   & 10$^5$--$10^5$       & 10$^2$ -- 10$^3$	\\
  LMC, SMC, IC\,10     & 8000 -- $10^5$ & --- 		      & ---			\\\hline
  \label{K99-results}
\end{tabular}
\tablefoot{For the outer regions of M\,33 and for the LMC, SMC, and IC\,10, 
  the standard model fails.}
\end{table}

\subsection{Diagnostic diagram of \CII /FIR vs. CO/FIR}
\label{pdr-model}

In Figure~\ref{hailey-todo} we plot luminosities of \CII\ versus CO,
normalized with FIR luminosities. To this diagnostic diagram we added
observations of other galaxies and the corresponding lines of constant
FUV-fields and local volume densities of standard K99 PDR models with
a cloud optical extinction of $A_{\rm V}=10$\,mag for a solar
metallicity. For completeness, we show in the Appendix
\ref{sec-hailey-tir} \CII\ vs. CO, but normalized to the
total-infrared (TIR) luminosities.

\subsubsection{Observations}

\paragraph{M33.} The ISO/LWS \CII/FIR ratios observed on scales of
280\,pc in M\,33 vary between $\sim0.8$\% in the inner parts of M\,33,
rising to up to 3\% in the outer regions, as already presented in the
previous sections. The \CII/CO 1--0 ratios vary between 1000 and 41200
while the CO 1--0/FIR ratios vary between $4\,10^{-7}$ and
$8\,10^{-6}$.

On scales of 50\,pc ($12''$), the \CII/FIR ratio varies between 0.01\%
and 3\% over the $2'\times2'$ map of the BCLMP\,302 \HII\ region
\citet[][M11]{mookerjea2011}\footnote{To convert the TIR values used
  by M11 into FIR, we used the FIR/TIR ratios derived from the
  greybody fits (Sec.~\ref{FIR-obs}, Tab.~\ref{Intensities}). Two
  ISO/LWS positions, N49 and N50, lie in the region mapped by M11.  We
  took the average FIR/TIR value for the conversion.}. The bulk of
\CII/FIR ratios lie in the range $\sim$0.7\%-1\%.  The CO 1--0/FIR
ratios lie between $4\,10^{-7}$ and $8\,10^{-6}$.  Interestingly, the
PACS observations of the $2'\times2'$ BCLMP\,302 region at 2.1\,kpc
radial distance cover the same range of \CII/FIR and CO 1--0/FIR
ratios as found with ISO/LWS along the entire major axis of M\,33 out
to 8\,kpc.


\paragraph{Other galaxies.} The \CII/CO ratios found in the inner
parts of M33 lie in the same range of values found in the bulk of the
normal galaxies and ULIRGs shown here. The outer regions of M\,33 show
higher values, similar to those found in other low-metallicity systems.

The ISO/LWS \CII/FIR ratios of M\,33 are higher than in normal
galaxies which only exhibit \CII/FIR ratios of up to 0.4\%.  The low
metallicity galaxies LMC, SMC, IC\,10 show high \CII/FIR ratios,
comparable to those found in M\,33, as already discussed above.

Local ULIRGs show CO/FIR ratios of less than $10^{-6}$, while normal
galaxies show higher ratios of up to $5\,10^{-6}$.  M33 shows slightly
higher peak ratios of up to $8\,10^{-6}$.

\subsubsection{PDR model results.} 

\paragraph{Observations consistent with the standard model.} Most of
the \CII/FIR and CO/FIR ratios observed in the inner disk of M\,33 lie
in the parameter space of local densities and FUV fields spanned by
the standard PDR of K99 (solar metallicity $Z=1.0$, $A_V=10\,$mag,
Fig.\,\ref{hailey-todo}). These ratios indicate densities of less than
$\sim10^4$\,\cmcub\ and FUV fields of less than $G_0\sim10^3$ in units
of the local interstellar value. Some of the ratios observed in the
inner galaxy and all ratios observed in the outer disk, are however
not consistent with this standard model, as discussed further below.

In the literature, the observed CO intensities have sometimes been
multiplied by a factor of 2 for the comparison with the Kaufman PDR
model.  \citet{hailey-dunsheath2010} argue that CO 1--0 is optically
thick stemming only from the frong-side of FUV illuminated clouds
while the FIR emission is in general optically thin stemming from the
front and the backside of clouds. Here, we do not apply any factor but
rather use the observed values argueing that the optical depth of the
galactic CO emission on scales of 280\,pc will be reduced due to the
velocity dispersion by turbulence and large-scale gradients, and that
therefore an ad-hoc factor of 2 does not seem appropriate. 


In the framework of the standard PDR model, normal galaxies and
especially the ULIRGs, tend to be consistent with higher densities of
up to a few times $10^6$\,cm$^{-3}$ than typically found in M\,33. The
latter are also consistent with higher FUV fields of up to $\sim10^4$.

Table~\ref{K99-results} summarizes the range of values returned by the
standard K99 PDR model for the whole data sample shown in
Figure~\ref{hailey-todo}.  Normal galaxy and starburst nuclei points
with extreme high \CII/CO values were ignored here.  


\paragraph{Observations inconsistent with the standard model.}  
All ISO/LWS \CII/FIR and CO/FIR ratios of the outer disk of M\,33 are
inconsistent with the standard PDR model for any density and FUV
field. In general, \CII/FIR ratios above 2\% cannot be reproduced by
the standard model for CO/FIR ratios of less than $10^{-5}$.  For
lower CO/FIR ratios, the upper limit of \CII/FIR which can be modeled,
drops. For instance, at CO/FIR$\sim10^{-6}$, the upper limit of
\CII/FIR lies near $0.5\%$.


The scatter of \CII/FIR and CO/FIR data seen in M\,33 on the large
scales sampled by the ISO/LWS beam is similar to the scatter seen on
small scales of 50\,pc in the small sub-region BCLMP\,302.  However,
variations of metallicities are not expected on such small scales. The
observed scatter of \CII/FIR and CO/FIR ratio must therefore reflect
the variation of other properties of the emitting gas and dust.

\begin{figure}
\centering
\includegraphics[angle=0,scale=0.4]{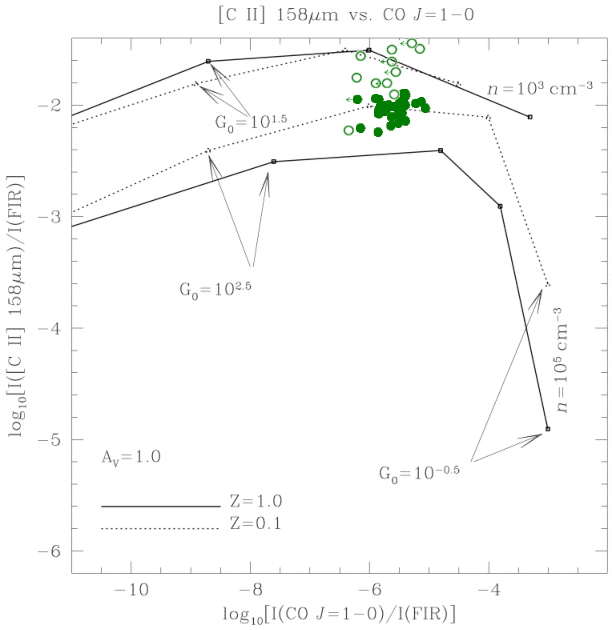}
\caption{ \CII\ versus CO, normalized with the FIR continuum
  (Figure\,18 of K99). Black solid and dotted lines indicate lines of
  constant density for metallicities of $Z=1.0$ and $Z=0.1$,
  respectivey, from the K99 PDR model with $A_{\rm {V}}=1$\,mag. Big
  green filled circles show ISO/LWS data of the inner S1, N1 regions
  of M\,33 while open circles show data of the outer S2, N2 regions. }
\label{hailey-av1mag}
\end{figure}

As shown in Figures\,\ref{hailey-todo} and \ref{hailey-av1mag}, K99
PDR models of low optical extinctions allow to reproduce the high
\CII/FIR ratios observed predominantly in the outer disk of M\,33.
Assuming a total column density corresponding to an $A_{\rm{V}}$ of
only 1\,mag and solar-metallicity, the ratios observed in the outer
disk are consistent with $n\sim10^3$\,\cmcub\ and $G_0\sim1$ while the
ratios observed in the inner disk are consistent with somewhat higher
densities and FUV fields of $n\sim10^4$\,\cmcub\ and $G_0\sim10$
(Fig.\,\ref{hailey-av1mag}). The ratio of $G_0/n$ is about
$10^{-3}$\,cm$^3$ in both cases.

Subsolar metallicities do not need to be invoked to explain these data
sets, as the curves for models with subsolar-metallicity $Z=0.1$ are
similar to the curves for $Z=1.0$. This also shows that \CII\ is
relatively insensitive to changes in the metallicity.  When \CII\
dominates gas-cooling, the models indicate that the gas temperature of
the PDR surface layer emitting \CII\ adjusts in a way that the
emergent \CII\ flux equals the FUV flux times the gas heating
efficiency \citep[][]{kaufman2006}.

As the observed metallicity gradient of M\,33 is shallow
\citep{magrini2010}, we do not suspect that the average optical
extinction of clouds changes abruptly from about 10\,mag to about
1\,mag between the inner and outer disk of M\,33. We suggest instead,
that the overall low-metallicity environment of M\,33 is composed of
clouds of in general low optical extinctions of typically only 1 or a
few magnitudes. 

%
In this scenario, the rather abrupt increase of heating efficiency
beyond about 4.5\,kpc radial distance is caused by a drop of average
local densities of the molecular gas and a decrease of FUV field
strengths.  

This drop of the FUV along the major axis of M\,33 is already
indicated by the radially averaged extinction corrected FUV fluxes
\citep[cf.\,Fig.\,3 in][]{verley2009}.
%
\citet{verley2007} did IR photometry of 515 compact sources. The best
models to reproduce the extinction seen in these \HII\ regions are the
ones with $A_{\rm V}<10\,$mag.


The change of density from $10^4$ in the inner disk to $10^3$ in
the outer disk, would imply that PDRs in the outer disk are typically
about a factor 10 larger in size than in the inner disk:

\begin{equation}
\frac{r_{\rm outer}}{r_{\rm inner}} = 
\frac{N({\rm H}_2)_{\rm outer}/n({\rm H}_2)_{\rm outer}}
     {N({\rm H}_2)_{\rm inner}/n({\rm H}_2)_{\rm inner}}
\sim 10.
\end{equation}


We estimated the FUV field in Habing units~\citep{habing1968} also
directly from the FIR continuum, using G$_0$=4$\pi$I$_{FIR}$/1.6$\cdot
10^{-3}$ in erg\,s$^{-1}$\,cm$^{-2}$ \citep{mookerjea2011,
  kaufman1999}, obtaining G$_0$ = 28 for the nucleus and $G_0=1.3$ for
the N66 position in the outer disk of M\,33.  The low FUV field found
in the outer disk is consistent with the radiation field predicted by
the K99 PDR model from the \CII/FIR and CO/FIR ratios for low
extinctions.


For a fixed CO/FIR ratio, and a given density and metallicity, the
modelled \CII/FIR decreases with increasing optical extinction
\citep[Figs.\,16-18 in][]{kaufman1999}.  For example, at
$\log$(CO/FIR)=6, $\log$(\CII/FIR) decreases by a factor $\sim5$ from
$-1.5$ at 1\,mag to $-2.2$ at 10\,mag.
%
In clouds of low optical extinctions, FUV photons penetrate deeper
into the cloud. For spherical clouds, the smaller CO core is
surrounded by a larger \CII\ emitting region, leading to enhanced
\CII/CO ratios \citep[cf.][]{bolatto1999,roellig2006}. The high
observed \CII/FIR and \CII/CO ratios are best explained by clouds of
low columns (optical extinctions) leading to an increase of the \CII\
layer relative to the total gas traced by CO, and also relative to the
total dust traced by the FIR continuum.  We propose that reduced
volume densities and the geometrical dilution of the FUV field when
photons can enter more deeply, are of secondary importance only.
%
%

The low optical extinctions are consistent with a reduced dust-to-gas
ratio in the low metallicity environment of M\,33.  As discussed by
\citet{israel1996} and \citet{israel-maloney2011} using data of the
LMC, the ratio of total gas column density over optical extinction,
$N_{\rm H}$/$A_{\rm V}$, increases in low metallicity environments.

\begin{figure}[h]
   \centering
     \includegraphics[scale=0.30,angle=0]{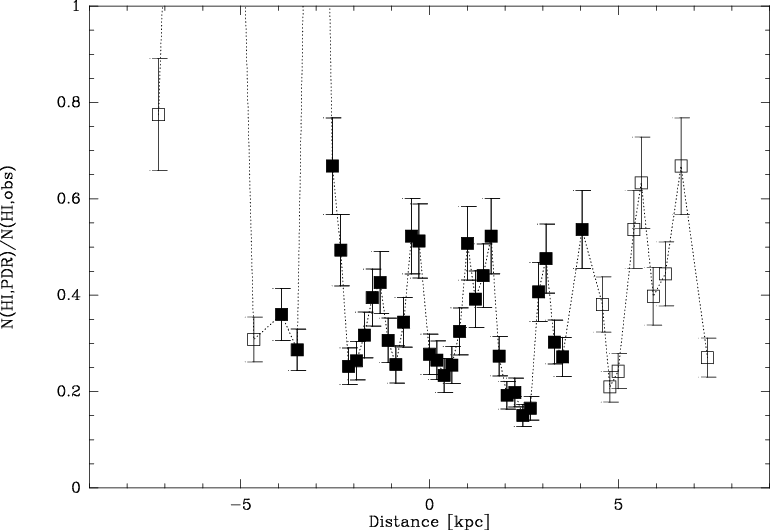}\\
     \caption{ Fraction of \HI\ column density stemming from PDRs.
       Errorbars only show the 15\% observational error of \HI\
       intensities. Two positions in the south for which
         $N$(\HI,PDR)$>N$(\HI,obs) are not shown.}
   \label{fig-npdr}             
\end{figure}

\subsection{\HI\ emission from PDRs}
\label{app-hipdr}

In the following, we will assume that all observed \HI\ emission stems
from the cold neutral and atomic medium (CNM) and from
photon-dominated regions. Before discussing a possible contribution of
\CII\ emission from the CNM, we first need to study the \HI\ fraction
stemming from PDRs. At the surface of photon dominated regions,
molecular hydrogen is photodissociated and a layer of atomic hydrogen
is formed. The column from \HI\ is a function of impinging FUV field
over the volume density of the molecular gas, $G_0/n$, as shown e.g.
by \citet[][]{sternberg1988}.  \citet[][]{stacey1991} studied the
contribution of \HI\ emission from PDRs to the total hydrogen column
in a sample of galaxies. With the typical densities and FUV fields
derived above for M\,33, we find a typical $G_0/n$ ratio of
$10^{-3}$\,cm$^{3}$ in the inner and outer disk of M\,33. Next we use
Eq.\,3 of \citet{heiner2011}:

  \begin{equation}
    N({\rm HI,PDR}) =
    \frac{7.8\,10^{20}}{D}
    \ln\Bigl(1+\frac{106\,G_0}{n}\,D^{-0.5}\Bigr) \,\,\rm{cm}^{-2}
  \end{equation}

  with a dust-to-gas ratio $D$ normalized to the solar neighborhoud
  value of $(12+\log({\rm O/H}))-8.69$ (Eq.\,2 of \citet{heiner2011})
  and a constant oxygen abundance of $12+\log({\rm O/H})=8.27$
  \citep[see references in][]{buchbender2013}.

  This leads to a constant \HI\ column density stemming from PDRs of
  $3.25\,10^{20}$\,cm$^{-2}$.

  In Figure\,\ref{fig-npdr}, we compare the \HI\ column density from
  PDRs with the observed beam-averaged \HI\ column density, assuming
  that the PDRs fill the beam. We did not attempt detailed PDR
  modeling to derive the beam filling factor, e.g. by comparing the
  observed FIR continuum with the best fitting FUV field from models.
  \citet{stacey1991} derive a typical beam filling factor of 0.3 for
  the sample of galaxies they studied. In M\,33, the estimated
  fraction of \HI\ column densities stemming from PDRs stays between
  15\% and 70\% for all data between 2\,kpc radial distance in the
  south and 7.3\,kpc in the north. Two positions further south exhibit
  very low \HI\ columns which are lower than the estimated \HI\ column
  from PDRs. Their $N$(\HI,PDR)/$N$(\HI,obs) ratios rise above 1 to
  values of up to 4, indicating that for these regions the underlying
  assumptions of $G_0$, $n$, and the beam filling factor are not
  valid.

\begin{figure}[h]
   \centering
   \includegraphics[scale=0.3,angle=0]{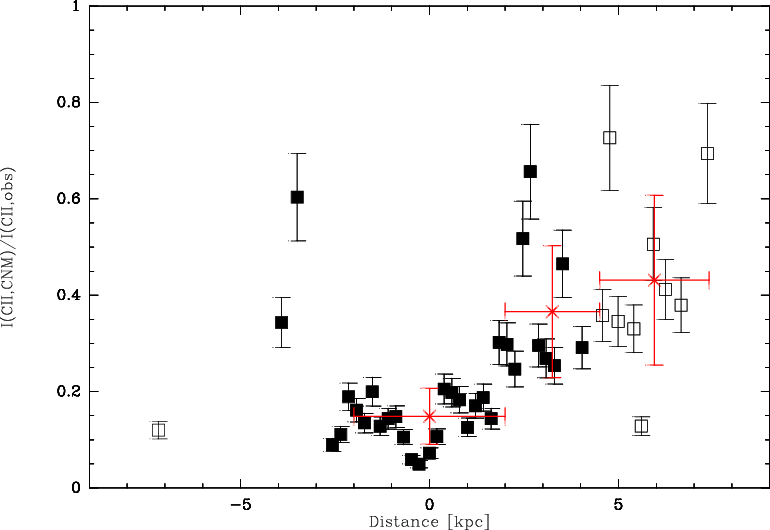}\\
   \caption{ Fraction of the observed \CII\ emission stemming from the
     cold neutral medium for $T$=80\,K and $n$=100\,\cmcub, after
     correction of the observed \HI\ for the contribution from PDRs.
     Squares mark the individual data along the major axis.  Their
     errorbars only include the 15\% observational error of the \HI\
     intensities. Crosses in red mark the fractions averaged over the
     inner 2\,kpc, the northern 2 to 4.5\,kpc, and the outer, northern
     disk between 4.5 and 7.4\,kpc.  Corresponding errorbars show the
     standard deviation of the individual data and the radius interval
     over which the values were binned. }
   \label{fig-cnm}             
\end{figure}

\subsection{\CII\ from the atomic medium}

Here, we investigate which fraction of the observed \CII\ emission
could stem from the atomic cold neutral medium (CNM) in M\,33.

As described in Section\,\ref{radial}, the emission of \HI\ drops only
weakly towards the outskirts, while the CO emission drops steeply
(Fig.\,\ref{distribution}).  In the southern part of the strip, the
\HI/CO rises continuously from a minimum in the center towards the
outskirts at about 5\,kpc radial distance (Fig.\,\ref{ciico-dist}).
In the North, on the other hand, the \HI/CO shows an abrupt increase
at about 4.5\,kpc radial distance, i.e. at the same distance where the
\CII/FIR ratio rises to values of $\sim3$\% (Fig.\,\ref{pehe-dist}).
The relative contribution of the atomic gas tracing the CNM to the
total gas emission (CO$+$\HI) rises with galacto-centric distance.
Also, the \HI/\CII\ intensity ratio rises strongly in the northern
part of the strip over almost an order of magnitude
(Fig.\,\ref{ciihi-dist}). This is because the \CII\ emission drops
more steeply with radius than the \HI. In the north, the observed
\CII\ emission has dropped by a factor 27 relative to its value at the
nucleus while \HI\ has dropped only by 40\%. Any contribution of the
CNM to the \CII\ emission relative to the contribution from PDRs
should be most prominent in the outskirts.

We note that the correlation between \HI\ and \CII\ emission is rather
weak throughout the strip, and also in the outskirts
(cf.\,Fig.\,\ref{correlaciones}).  \citet[][]{mookerjea2011} and
\citet{braine2012} studied \CII\ HIFI spectra in two regions of
$\sim1'-2'$ along the major axis of M\,33 at radial distances of 2.1
and 3.3\,kpc.  In both sources, the \HI\ lines are shifted
systematically relative to the \CII\ lines by about 5\,kms$^{-1}$. In
addition, \HI\ linewidths are about 30\% broader than those of \CII.
These findings indicate that \CII\ emission, at least in these two
regions of the inner galaxy, does not simply trace only the atomic
medium.

To quantify a possible contribution of the cold neutral medium (CNM)
to the \CII\ emission, we use the \HI\ line intensity, corrected for
the contribution from PDRs, to derive the intensity of the \CII\ line
\citep[cf.][]{crawford1985, madden1993, madden1997, langer2010}. Here,
we assume optically thin \CII\ and \HI\ emission and beamfilling
factors of unity. We estimate the \CII\ intensities stemming from the
remaining fraction of the atomic gas via:

\begin{eqnarray}
   I'({\rm HI})   & = & 3.35\,10^{14}\,I({\rm HI}) \\
   N({\rm HI})   & = & 1.82\,10^{18}\,I'({\rm HI}) \\
 N({\rm HI,CNM}) & = & N({\rm HI})-N({\rm HI,PDR}) \\
 N({\rm{C}^+, \rm{CNM}}) & = & X_{{\rm C}^+}\,N({\rm HI, CNM}) \\
 I({\rm{C}^+, \rm{CNM}}) & = & 2.35\,10^{-21}  N({\rm{C}^+}) \times \\
 & & \Bigl(
    \frac
    {2\,\exp(-\Delta E/T)}
    {1+2\,\exp(-\Delta E/T)+(n_{\rm cr}/n)}
 \Bigr)
\end{eqnarray}


with intensities $I$ in erg\,s$^{-1}$\,cm$^{-2}$\,sr$^{-1}$, and $I'$
in K\,km\,s$^{-1}$, and column densities in cm$^{-2}$, the energy of
the \CII\ $^2P_{3/2}$ state above ground $\Delta E=h\nu/k=91.3\,$K,
and the critical density for collisions with hydrogen atoms and/or
molecules of $n_{\rm cr}=3\,10^{3}$\,cm$^{-3}$ \citep[cf. discussion
and references in][]{langer2010}. The fractional abundance $X$(C$^+$)
of C$^+$ in the gas with respect to \HI\ lies in the range 1.4 to
1.8\,$10^{-4}$ in the local ISM of the Milky Way \citep{sofia1997}.
For the average low metallicity environment of M\,33, we assume
$X$(C$^+$)=0.6\,10$^{-4}$ \citep{magrini2010, henry2000}. In addition,
we assume diffuse atomic clouds of $n=100$\,cm$^{-3}$ and $T=80\,$\,K
corresponding to a pressure of about 8000\,K\,cm$^{-3}$ typical for
the diffuse atomic medium in the Milky Way
\citep[e.g.][]{wolfire1995,dickey2000}.  Higher temperatures and
densities would increase the \CII\ intensities but quickly become
inconsistent with the generally accepted values for the pressure of
the Galactic atomic ISM \citep[cf. discussion in Sec.\,4.2.1.
of][]{madden1993}.




%
%

Figure\,\ref{fig-cnm} shows the estimated fraction of the observed
\CII\ intensity which stems from the CNM. As expected from the
observed ratio of \HI\ over \CII\ intensities, the fraction is lowest
in the nucleus and in general rises towards the outskirts.  Towards
the southern half, the CNM fraction shows strong variability
reflecting the variations of \HI\ emission on small scales
(cf.\,Fig.\,\ref{distribution}). However, towards the north, the
fraction rises in general from $20\%\pm6\%$ in the inner 2\,kpc, to
$37\%\pm14\%$ between 2 and 4.5\,kpc, to $43\%\pm18\%$ between 4.5 and
7.4\,kpc radial distance.

The rise of \CII/FIR by a factor of more than 3 with distance may
therefore be partially explained by a rising contribution of the CNM
to the \CII\ emission while the \CII/FIR ratio of PDRs rises only
moderately with only somewhat lowered values of $A_V$, $n$, $G_0$
relative to the PDR models for the inner galaxy of $A_V\sim10$\,mag,
$n\sim10^4$\,cm$^{-3}$, $G_0\sim10$. A better knowledge of the
temperature and density of the CNM is needed to derive more
quantitative conclusions.

%
%


Here, we conclude that PDRs dominate in the inner parts of M\,33 while
both, the CNM and PDRs, can explain the \CII\ emission observed in the
outskirts.

%

\subsection{The importance of the \OI\ 63$\mu$m line}

A more complete study of the major gas cooling lines of the ISM would
need to include the \OI\ 63$\mu$m line as a tracer of high densities
in PDRs. \citet{higdon2003} detected this line at scales of 280\,pc in
the nucleus and towards five \HII\ regions in M\,33, finding \CII/\OI\
ratios between 1.6 and 2.2.  Towards the \HII\ region BCLMP\,302 and
on scales of 50\,pc, the ratio varies betwen 2.5 and 10
\citep{mookerjea2011}. Further Herschel/PACS and HIFI observations of
several regions along the major axis of M\,33 have been conducted in
the framework of the \hermes\ project, to obtain a more complete
dataset
which will allow e.g.  to derive more accurate estimates of the
photoelectric heating efficiency and its variation on small scales, as
also it will allow to determine gas properties ($A_V, n, T$) using PDR
models.

\section{Summary and conclusions}

In this paper we have studied ISO/LWS \CII\ emission along the major
axis of M\,33 at scales of 280\,pc. Stacking of few adjacent positions
allowed to detect \CII\ between galacto-centric distances of
$\sim6\,$kpc in the South to $\sim6.7$\,kpc in the North.

The radial distributions of the \CII/CO and \HI/CO ratios show an
increase from the inner regions to the outer regions. While the \HI/CO
ratio has a minimum near the nucleus, the \CII/CO ratio shows a
minimum at about $10'$ to the North, corresponding to the global
maximum of CO emission along the major axis.

The radial distribution of \CII\ shows a strong correlation with star
formation tracers, the FIR continuum and H$\alpha$ emission, closely
following the spiral arm structure. The correlation becomes weaker in
the outskirts of M\,33 beyond 4.5\,kpc radial distance to the nucleus.

The \CII/FIR ratio remains constant at $\sim0.8$\% in the central
4.5\,kpc radius and then increases rapidly to values of up to
3\%.  The \CII/FIR ratio is a strong function of the FIR luminosity.
The highest \CII/FIR ratios are found for the lowest FIR luminosities.
%
Other low metallicity systems (LMC, SMC, IC\,10) show similar \CII/FIR
ratios as M\,33 which are higher than those found in the Milky Way and
in normal galaxies.
	

The variation of \CII/FIR and CO/FIR ratios found in a small
$2'\times2'$ \HII\ region of M\,33 on small scales of 50\,pc
\citep{mookerjea2011} is very similar to the variation found with the
ISO/LWS observations. This is another indication that metallicity
variations play a minor role in determining the \CII/FIR ratio.

Diagnostic plots of \CII/FIR vs. CO/FIR or FIR/CO vs. \CII/CO show
smooth transitions from normal galaxies to the positions in the inner
4.5\,kpc radius of M\,33, to the outer parts of M\,33 together
with the other low-metallicity systems. Observations of the inner disk
of M\,33 serve as bridge between normal galaxies exhibiting low
\CII/FIR and low \CII/CO ratios to the low-metallicity systems showing
high \CII/FIR ratios and also high \CII/CO ratios.

The relative lack of dust shielding in the low metallicity systems
enhances CO photo-dissociation, and, hence, increases the \CII/CO
ratios. The standard K99 PDR models of 10\,mag optical extinction fail
to reproduce the high end of the observed \CII/FIR and \CII/CO ratios.
However, models of low optical extinction expected in the
low-metallicity environment of M\,33 do reproduce the observed ratios
for a constant metallicity. In the framework of models of
$A_V=1$\,mag, the variation of observed \CII/FIR ratios is caused by
variations of the local densities dropping from about
$10^4$\,cm$^{-3}$ in the inner disk to $10^3$\,cm$^{-3}$ in the outer
parts. In addition, FUV field strengths reach $\sim10$ times the
average interstellar radiation field in the inner disk while dropping
to only a few in the outskirts. High FUV fields tend to lower the
photoelectric heating efficiency as grains and PAHs become positively
charged \citep{okada2013}.  In contrast, the high \CII/FIR ratios
observed in low metallicity galaxies can be caused by geometric
dilution of low FUV fields in clouds of normal densities where Carbon
stays ionized.

In addition to PDRs, the atomic cold neutral medium (CNM) traced by
\HI\ may contribute to the observed \CII\ emission. Therefore, the
above scenario may need to be revised. We conclude that the CNM,
corrected for the \HI\ emission from PDRs, contributes $\sim15\%$ to
the \CII\ observed in the inner radius of 2\,kpc.  However, the CNM
may contribute about 40\% of the observed \CII\ in the outer, northern
disk where \HI\ emission is much stronger relative to \CII.



\begin{acknowledgements} 

  We thank S.\,Hailey-Dunsheath and J.\,Gracia-Carpio for providing us
  data shown in Figure\,\ref{hailey-todo}. We thank an anonymous
  referee for insightful comments.

\end{acknowledgements}   



\bibliography{kramer-lws} 
\bibliographystyle{aa} 

\begin{appendix} 

\section{List of positions}

The Tables~\ref{PosSur} and~\ref{PosNor} list the 77 positions
observed with ISO/LWS, their galacto-centric distance, and whether
\CII\ was detected.

\begin{table*}
  \caption{ISO/LWS positions observed in \CII\ along the southern part of
    the major axis of M\,33. }
		\centering                          
		\begin{tabular}{c c c c c c}        
		\hline\hline                 
Pos ID	&	RA 		& 	DEC 		&  Ang. Dis [$'$] &	Lin. Dis [kpc] & \CII\ detection\,?\\    
 (1) & (2) & (3) & (4) & (5) & (6) \\
		\hline     
S1	&	01:32:57	&	+30:10:06	&	-32.70	&	-7.99	&	S1-S8	\\
S2	&	01:32:58	&	+30:10:52	&	-31.80	&	-7.77	&	S1-S8	\\
S3	&	01:32:59	&	+30:11:38	&	-31.00	&	-7.57	&	S1-S8	\\
S4	&	01:33:01	&	+30:12:24	&	-30.10	&	-7.35	&	S1-S8	\\
S5	&	01:33:02	&	+30:13:10	&	-29.30	&	-7.16	&	S1-S8	\\
S6	&	01:33:03	&	+30:13:56	&	-28.40	&	-6.94	&	S1-S8	\\
S7	&	01:33:05	&	+30:14:42	&	-27.50	&	-6.72	&	S1-S8	\\
S8	&	01:33:06	&	+30:15:28	&	-26.70	&	-6.52	&	S1-S8	\\
S9	&	01:33:07	&	+30:16:14	&	-25.80	&	-6.30	&	S9-S14	\\
S10	&	01:33:08	&	+30:17:00	&	-25.00	&	-6.11	&	S9-S14	\\
S11	&	01:33:10	&	+30:17:46	&	-24.10	&	-5.89	&	S9-S14	\\
S12	&	01:33:11	&	+30:18:32	&	-23.30	&	-5.69	&	S9-S14	\\
S13	&	01:33:12	&	+30:19:18	&	-22.40	&	-5.47	&	S9-S14	\\
S14	&	01:33:14	&	+30:20:04	&	-21.60	&	-5.28	&	S9-S14	\\
S15	&	01:33:15	&	+30:20:50	&	-20.70	&	-5.06	&	S15-S19	\\
S16	&	01:33:16	&	+30:21:36	&	-19.90	&	-4.86	&	S15-S19	\\
S17	&	01:33:18	&	+30:22:22	&	-19.00	&	-4.64	&	S15-S19	\\
S18	&	01:33:19	&	+30:23:08	&	-18.10	&	-4.42	&	S15-S19 \\
S19	&	01:33:20	&	+30:23:54	&	-17.30	&	-4.23	&	S15-S19 \\
S20	&	01:33:24	&	+30:24:40	&	-16.40	&	-4.01	&	S20-S21	\\
S21	&	01:33:25	&	+30:25:27	&	-15.60	&	-3.81	&	S20-S21	\\
S22	&	01:33:27	&	+30:26:13	&	-14.70	&	-3.59	&	S22-S23	\\
S23	&	01:33:28	&	+30:26:59	&	-13.90	&	-3.40	&	S22-S23	\\
S24	&	01:33:30	&	+30:27:45	&	-13.00	&	-3.18	&	S24-S26	\\
S25	&	01:33:31	&	+30:28:31	&	-12.20	&	-2.98	&	S24-S26	\\
S26	&	01:33:33	&	+30:29:17	&	-11.30	&	-2.76	&	S24-S26	\\
S27	&	01:33:34	&	+30:30:03	&	-10.50	&	-2.57	&	X	\\
S28	&	01:33:36	&	+30:30:49	&	-9.60	&	-2.35	&	X	\\
S29	&	01:33:37	&	+30:31:35	&	-8.75	&	-2.14	&	X	\\
S30	&	01:33:39	&	+30:32:21	&	-7.89	&	-1.93	&	X	\\
S31	&	01:33:40	&	+30:33:07	&	-7.04	&	-1.72	&	X	\\
S32	&	01:33:42	&	+30:33:53	&	-6.19	&	-1.51	&	X	\\
S33	&	01:33:43	&	+30:34:39	&	-5.33	&	-1.30	&	X	\\
S34	&	01:33:45	&	+30:35:25	&	-4.48	&	-1.09	&	X	\\
S35	&	01:33:46	&	+30:36:11	&	-3.63	&	-0.89	&	X	\\
S36	&	01:33:48	&	+30:36:57	&	-2.79	&	-0.68	&	X	\\
S37	&	01:33:49	&	+30:37:43	&	-1.95	&	-0.48	&	X	\\
S38	&	01:33:51	&	+30:38:29	&	-1.14	&	-0.28	&	X	\\
\hline
\label{PosSur}
\end{tabular}
\tablefoot{Columns (4) and (5) list galacto-centric
  distances. Negative values indicate southern positions. Column (6)
  indicates whether \CII\ emission was detected above $3\sigma$ at a
  given position (marked by ``X'') or were averaged over a given range
  of positions.
}
\end{table*}

\begin{table*}
  \caption{ISO/LWS positions observed in \CII\ along the northern part of
    the major axis of M\,33.}  
\centering                          
\begin{tabular}{c c c c c c}        
\hline\hline                 
Pos ID	&	RA (J2000)	& 	DEC (J2000)	&  Ang. Dis [$'$] &	Lin. Dis [kpc] & \CII\ Detection ?\\    
 (1) & (2) & (3) & (4) & (5) & (6) \\
\hline                        
39 (Nucleus)	&	01:33:51	&	+30:39:37	&	0.00	&	0.00	&	X	\\
N40	&	01:33:54	&	+30:40:01	&	.80	&	0.20	&	X	\\
N41	&	01:33:55	&	+30:40:47	&	1.58	&	0.39	&	X	\\
N42	&	01:33:57	&	+30:41:33	&	2.42	&	0.59	&	X	\\
N43	&	01:33:58	&	+30:42:19	&	3.26	&	0.80	&	X	\\
N44	&	01:34:00	&	+30:43:05	&	4.11	&	1.00	&	X	\\
N45	&	01:34:01	&	+30:43:51	&	4.96	&	1.21	&	X	\\
N46	&	01:34:03	&	+30:44:37	&	5.81	&	1.42	&	X	\\
N47	&	01:34:04	&	+30:45:23	&	6.67	&	1.63	&	X	\\
N48	&	01:34:06	&	+30:46:09	&	7.52	&	1.84	&	X	\\
N49	&	01:34:07	&	+30:46:55	&	8.38	&	2.05	&	X	\\
N50	&	01:34:09	&	+30:47:41	&	9.23	&	2.26	&	X	\\
N51	&	01:34:10	&	+30:48:27	&	10.1	&	2.47	&	X	\\
N52	&	01:34:12	&	+30:49:13	&	10.9	&	2.66	&	X	\\
N53	&	01:34:13	&	+30:49:59	&	11.8	&	2.88	&	X	\\
N54	&	01:34:15	&	+30:50:45	&	12.6	&	3.08	&	X	\\
N55	&	01:34:16	&	+30:51:31	&	13.5	&	3.30	&	X	\\
N56	&	01:34:18	&	+30:52:17	&	14.4	&	3.52	&	X	\\
N57	&	01:34:19	&	+30:53:03	&	15.2	&	3.71	&	N57-N60	\\
N58	&	01:34:21	&	+30:53:49	&	16.1	&	3.93	&	N57-N60	\\
N59	&	01:34:23	&	+30:54:35	&	16.9	&	4.13	&	N57-N60	\\
N60	&	01:34:24	&	+30:55:21	&	17.8	&	4.35	&	N57-N60	\\
N61	&	01:34:26	&	+30:56:07	&	18.7	&	4.57	&	X	\\
N62	&	01:34:27	&	+30:56:53	&	19.5	&	4.76	&	X	\\
N63	&	01:34:29	&	+30:57:39	&	20.4	&	4.98	&	X	\\
N64	&	01:34:30	&	+30:58:25	&	21.2	&	5.18	&	-	\\
N65	&	01:34:32	&	+30:59:11	&	22.1	&	5.40	&	X	\\
N66	&	01:34:33	&	+30:59:57	&	22.9	&	5.60	&	X	\\
N67	&	01:34:35	&	+31:00:43	&	23.8	&	5.82	&	N67-N68	\\
N68	&	01:34:36	&	+31:01:29	&	24.6	&	6.01	&	N67-N68	\\
N69	&	01:34:38	&	+31:02:15	&	25.5	&	6.23	&	X	\\
N70	&	01:34:39	&	+31:03:01	&	26.4	&	6.45	&	N70-N72	\\
N71	&	01:34:41	&	+31:03:47	&	27.2	&	6.65	&	N70-N72	\\
N72	&	01:34:42	&	+31:04:33	&	28.1	&	6.87	&	N70-N72	\\
N73	&	01:34:44	&	+31:05:19	&	28.9	&	7.06	&	N73-N77	\\
N74	&	01:34:45	&	+31:06:05	&	29.8	&	7.28	&	N73-N77	\\
N75	&	01:34:47	&	+31:06:51	&	30.6	&	7.48	&	N73-N77	\\
N76	&	01:34:48	&	+31:07:38	&	31.5	&	7.70	&	N73-N77	\\
N77	&	01:34:50	&	+31:08:24	&	32.3	&	7.89	&	N73-N77	\\
\label{PosNor}
\end{tabular}
\tablefoot{Columns (4) and (5) list galacto-centric distances. Column
(6) indicates whether \CII\ emission was detected above $3\sigma$ at a
given position (marked by ``X''), was averaged over a given range
of positions, or whether no emission was detected (marked by ``-'').
}
\end{table*}

\section{Intensities at the observed ISO/LWS positions }
\label{co10}

\begin{table*}
\caption{Integrated intensities at the positions with \CII\ detections.}             
\centering                          
\begin{tabular}{c c c c c c c c }        
\hline\hline                 
Pos. ID	&	\CII\	&	\HI&	H$\alpha$	&	CO(2-1)	&	CO 1--0	&	FIR	&	TIR/FIR	\\
\hline
S1-S8	&	\textbf{2.20e-06}	&	6.88e-13	&	\textbf{3.05e-07}	&	---	&	---	&	9.12e-05	&	2.52e+00	\\
S9-S14	&	1.28e-06	&	1.37e-13	&	\textbf{3.11e-07}	&	\textbf{1.18e-09}	&	\textbf{8.32e-11}	&	4.25e-05	&	2.50e+00	\\
S15-S19	&	1.26e-06	&	1.73e-12	&	8.68e-07	&	3.13e-09	&	2.39e-10	&	4.17e-05	&	2.47e+00	\\
S20-S21	&	4.70e-06	&	1.48e-12	&	3.75e-06	&	6.35e-09	&	5.15e-10	&	8.39e-04	&	1.54e+00	\\
S22-S23	&	3.75e-06	&	1.86e-12	&	1.20e-06	&	7.06e-09	&	5.87e-10	&	4.61e-04	&	1.89e+00	\\
S24-S26	&	2.69e-06	&	2.00e-13	&	9.64e-07	&	\textbf{1.70e-09}	&	\textbf{1.44e-10}	&	2.62e-04	&	2.32e+00	\\
S27	&	5.04e-06	&	7.98e-13	&	3.63e-06	&	3.10e-09	&	5.51e-10	&	4.82e-04	&	1.44e+00	\\
S28	&	8.38e-06	&	1.08e-12	&	7.35e-06	&	5.92e-09	&	1.04e-09	&	8.19e-04	&	1.48e+00	\\
S29	&	1.42e-05	&	2.11e-12	&	1.34e-05	&	1.69e-08	&	2.94e-09	&	1.73e-03	&	1.49e+00	\\
S30	&	1.57e-05	&	2.02e-12	&	1.61e-05	&	1.81e-08	&	3.11e-09	&	1.72e-03	&	1.73e+00	\\
S31	&	1.45e-05	&	1.68e-12	&	1.24e-05	&	1.28e-08	&	2.17e-09	&	1.62e-03	&	1.56e+00	\\
S32	&	6.97e-06	&	1.35e-12	&	7.45e-06	&	9.68e-09	&	1.63e-09	&	1.34e-03	&	1.38e+00	\\
S33	&	9.54e-06	&	1.25e-12	&	5.81e-06	&	8.88e-09	&	1.48e-09	&	1.16e-03	&	1.44e+00	\\
S34	&	1.43e-05	&	1.74e-12	&	9.92e-06	&	1.41e-08	&	2.32e-09	&	1.78e-03	&	1.42e+00	\\
S35	&	1.78e-05	&	2.08e-12	&	1.66e-05	&	2.75e-08	&	4.30e-09	&	2.00e-03	&	1.32e+00	\\
S36	&	1.64e-05	&	1.55e-12	&	1.41e-05	&	2.02e-08	&	3.15e-09	&	2.01e-03	&	1.39e+00	\\
S37	&	1.41e-05	&	1.02e-12	&	1.38e-05	&	1.50e-08	&	2.34e-09	&	1.85e-03	&	1.37e+00	\\
S38	&	1.75e-05	&	1.04e-12	&	1.90e-05	&	1.75e-08	&	2.74e-09	&	2.25e-03	&	1.34e+00	\\
39 (Nucleus)	&	3.27e-05	&	1.92e-12	&	2.77e-05	&	3.41e-08	&	5.34e-09	&	3.62e-03	&	1.34e+00	\\
N40	&	2.36e-05	&	2.01e-12	&	1.52e-05	&	2.85e-08	&	4.45e-09	&	2.78e-03	&	1.33e+00	\\
N41	&	1.45e-05	&	2.28e-12	&	1.03e-05	&	3.17e-08	&	4.96e-09	&	2.45e-03	&	1.39e+00	\\
N42	&	1.34e-05	&	2.09e-12	&	8.94e-06	&	3.72e-08	&	5.81e-09	&	2.16e-03	&	1.40e+00	\\
N43	&	1.03e-05	&	1.64e-12	&	9.89e-06	&	2.65e-08	&	4.14e-09	&	1.33e-03	&	1.69e+00	\\
N44	&	7.00e-06	&	1.05e-12	&	4.74e-06	&	1.38e-08	&	2.27e-09	&	8.56e-04	&	1.47e+00	\\
N45	&	8.27e-06	&	1.36e-12	&	5.99e-06	&	1.58e-08	&	2.62e-09	&	8.19e-04	&	1.73e+00	\\
N46	&	6.15e-06	&	1.21e-12	&	4.08e-06	&	1.11e-08	&	1.85e-09	&	6.41e-04	&	1.68e+00	\\
N47	&	5.77e-06	&	1.02e-12	&	2.85e-06	&	1.10e-08	&	1.86e-09	&	5.28e-04	&	1.63e+00	\\
N48	&	8.00e-06	&	1.95e-12	&	4.71e-06	&	1.83e-08	&	3.13e-09	&	8.83e-04	&	1.48e+00	\\
N49	&	1.28e-05	&	2.77e-12	&	1.56e-05	&	2.17e-08	&	3.75e-09	&	1.79e-03	&	1.43e+00	\\
N50	&	1.49e-05	&	2.69e-12	&	1.41e-05	&	2.39e-08	&	4.18e-09	&	1.30e-03	&	1.89e+00	\\
N51	&	9.91e-06	&	3.54e-12	&	5.56e-06	&	3.73e-08	&	6.59e-09	&	1.01e-03	&	1.96e+00	\\
N52	&	6.98e-06	&	3.22e-12	&	2.96e-06	&	3.51e-08	&	6.27e-09	&	8.13e-04	&	1.83e+00	\\
N53	&	4.48e-06	&	1.31e-12	&	1.37e-06	&	1.41e-08	&	2.56e-09	&	4.78e-04	&	1.42e+00	\\
N54	&	3.72e-06	&	1.12e-12	&	2.21e-06	&	6.24e-09	&	1.14e-09	&	3.22e-04	&	1.78e+00	\\
N55	&	8.24e-06	&	1.76e-12	&	1.43e-05	&	1.34e-08	&	2.48e-09	&	8.23e-04	&	1.72e+00	\\
N56	&	5.23e-06	&	1.96e-12	&	1.26e-05	&	1.50e-08	&	2.80e-09	&	7.96e-04	&	1.69e+00	\\
N57-N60	&	2.70e-06	&	9.94e-13	&	2.69e-06	&	2.27e-09	&	1.87e-10	&	5.12e-04	&	1.56e+00	\\
N61	&	4.13e-06	&	1.40e-12	&	2.26e-06	&	2.55e-09	&	5.07e-10	&	2.19e-04	&	1.90e+00	\\
N62	&	4.71e-06	&	2.54e-12	&	5.81e-06	&	4.41e-09	&	8.87e-10	&	4.09e-04	&	1.74e+00	\\
N63	&	8.22e-06	&	2.20e-12	&	8.17e-06	&	3.11e-09	&	6.35e-10	&	5.66e-04	&	1.46e+00	\\
N65	&	2.38e-06	&	9.94e-13	&	1.26e-06	&	3.57e-10	&	7.48e-11	&	1.47e-04	&	1.75e+00	\\
N66	&	4.10e-06	&	8.42e-13	&	8.15e-06	&	3.86e-10	&	9.97e-11	&	1.68e-04	&	1.71e+00	\\
N67-N68	&	2.72e-06	&	1.34e-12	&	4.44e-06	&	\textbf{4.26e-10}	&	\textbf{3.10e-10}	&	1.89e-04	&	2.70e+00	\\
N69	&	2.76e-06	&	1.20e-12	&	1.31e-06	&	\textbf{4.89e-09}	&	\textbf{3.45e-10}	&	8.09e-05	&	2.08e+00	\\
N70-N72	&	1.19e-06	&	7.98e-13	&	\textbf{3.19e-07}	&	\textbf{1.45e-09}	&	\textbf{1.01e-10}	&	5.06e-05	&	2.12e+00	\\
N73-N77	&	\textbf{3.53e-06}	&	1.97e-12	&	\textbf{5.25e-07}	&	\textbf{4.24e-10}	&	\textbf{2.83e-11}	&	8.51e-04	&	2.31e+00	\\

\label{Intensities}
\end{tabular}
\tablefoot{Intensities are given in erg
s$^{-1}$sr$^{-1}$cm$^{-2}$ at the angular resolution of the \CII\ data. 
Values in bold face are upper limits.}
\end{table*}

The IRAM 30m CO 2--1 data on the $T_{\rm{A}}^{*}$ scale were converted
into $T_{\rm mb}$ temperatures using the forward efficiency $F_{\rm
  eff} = 0.90$ and the main beam efficiency $B_{\rm eff} = 0.49$:
$T_{mb}$ = ($F_{\rm eff}/B_{\rm eff}$) $T_{A}^*$.  
The CO 2--1 map was first smoothed from the original $11''$ resolution
to 22$''$ using a Gaussian kernel.  To derive CO 1--0 intensities from
the CO 2--1 data, we used a linear function of the 2--1/CO 1--0 ratio
dropping from 0.8 in the nucleus of M\,33 to 0.5 at a galactocentric
distance of 8.5\,kpc \citep{gratier2010}. Next, we smoothed the map to
the ISO/LWS resolution and converted the data to intensities in erg
cm$^{-2}$s$^{-1}$sr$^{-1}$.


Both the \HI\ data and the H$\alpha$ data were also smoothed to the
LWS resolution. The original resolutions of the \HI\ data was $11''$
while we assumed an original pencil beam for the H$\alpha$ data.

Table~\ref{Intensities} lists for all positions and stacked areas the
resulting integrated intensities of \CII, \HI, H$\alpha$, CO 2--1,
1--0, the FIR continuum, and the TIR/FIR ratio.

\section{Sample \CII\ ISO/LWS spectra and SEDs} \label{sec-examples}

\begin{figure*}[ht!]
   \centering
   \subfloat[]{
        \label{spec1}         
        \includegraphics[scale=0.27,angle=0]{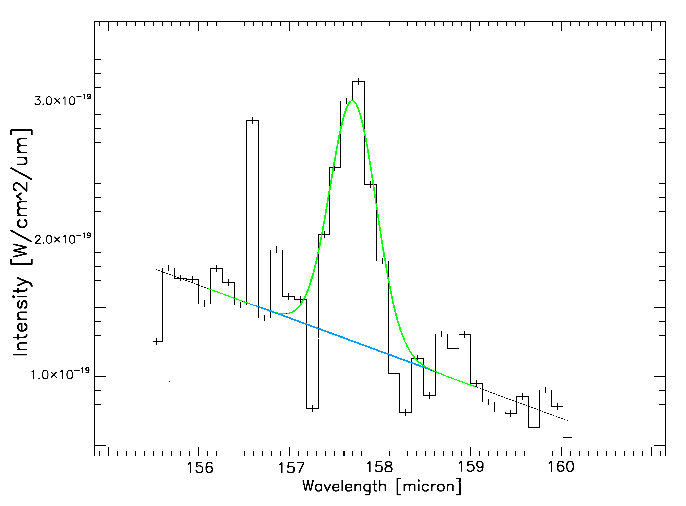}}
   \subfloat[]{
        \label{sed1}         
        \includegraphics[scale=0.33,angle=0]{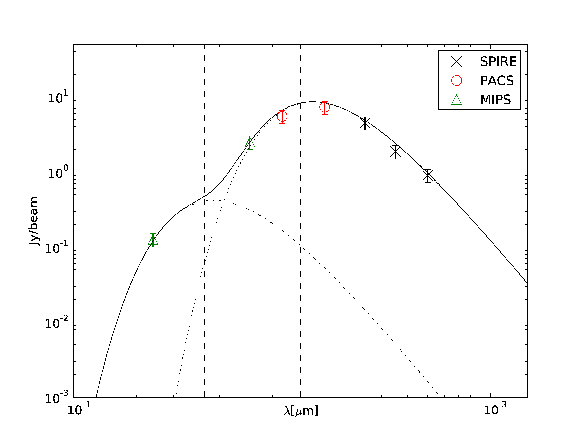}}\\

   \subfloat[]{
        \label{spec2}         
        \includegraphics[scale=0.27,angle=0]{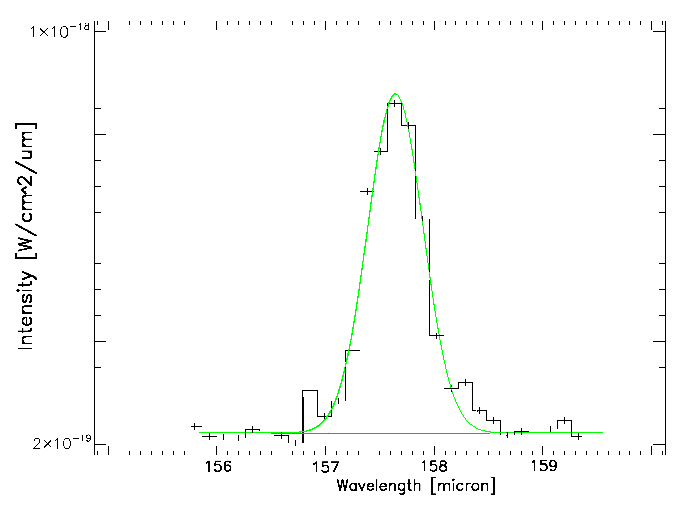}}
   \subfloat[]{
        \label{sed2}         
        \includegraphics[scale=0.33,angle=0]{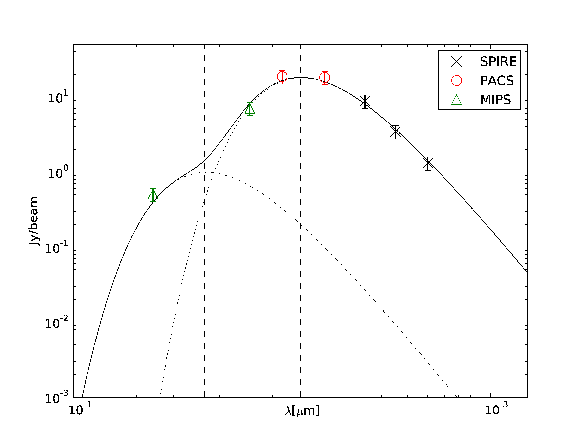}}\\
   \subfloat[]{
        \label{spec3}         
        \includegraphics[scale=0.27,angle=0]{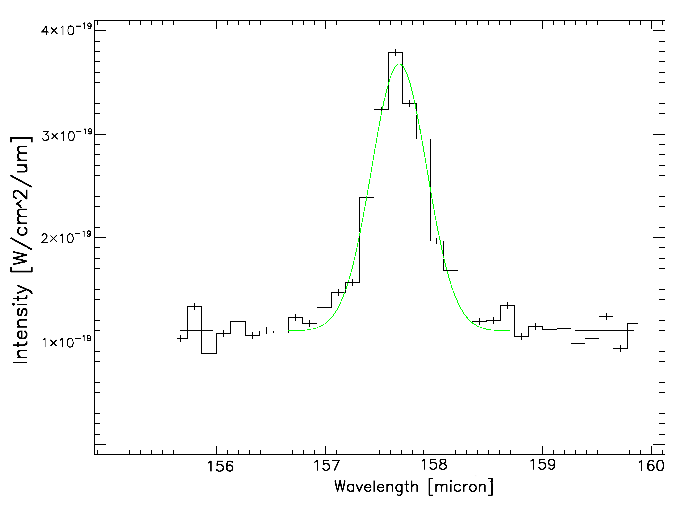}}
   \subfloat[]{
        \label{sed3}         
        \includegraphics[scale=0.33,angle=0]{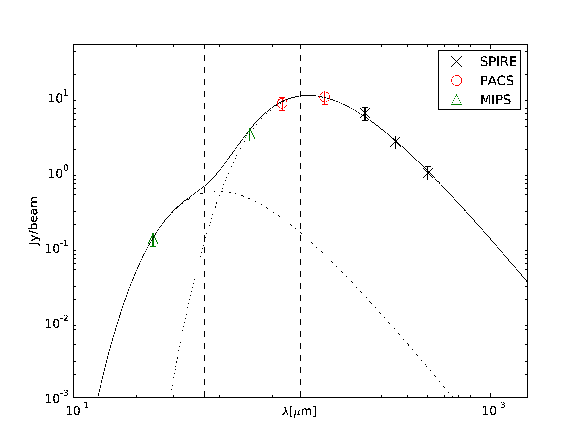}}\\
   \subfloat[]{
        \label{spec4}         
        \includegraphics[scale=0.27,angle=0]{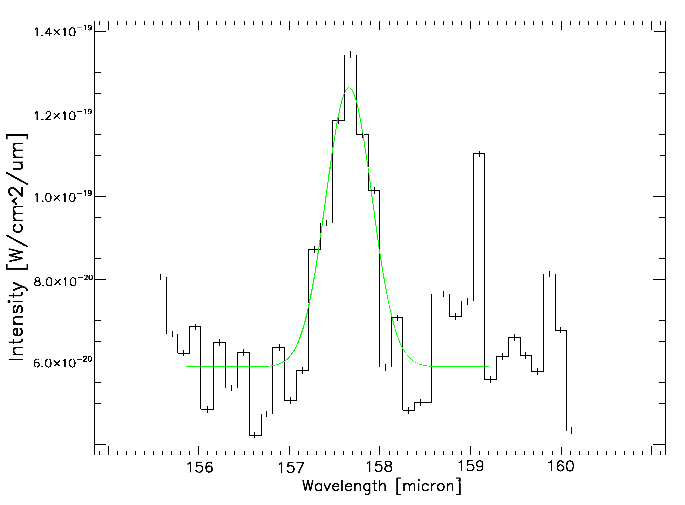}}
   \subfloat[]{
        \label{sed4}         
        \includegraphics[scale=0.33,angle=0]{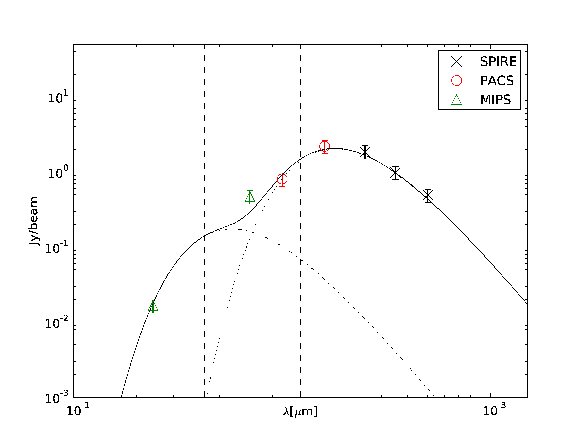}}\\

      \caption{Examples of some of the \CII\ ISO/LWS spectra and SEDs
        obtained in this work. The vertical dashed lines mark the
        integration interval to derive the FIR continuum, 42.5 $\mu$m
        and 122.5$\mu$m \citep{dale-helou2002}. {\bf a,b:} The first
        row shows the \CII\ spectrum and the SED of the inter-arm
        position S32.  {\bf c,d:} The second row shows the spectrum
        and SED from the nucleus of M\,33. {\bf e,f:} The third rows
        shows observations of the \HII\ region BCLMP\,302 (N49). {\bf
          g,h:} Finally, the last row shows the spectrum and the SED
        from one of the positions (N61) in the outer, southern N2
        region.  The SEDs show MIPS 24$\mu$m and 70$\mu$m, PACS
        100$\mu$m and 160$\mu$m and SPIRE 250$\mu$m, 350$\mu$m and
        500$\mu$m data. In the \CII\ spectra, green lines show
        Gaussian fits to the spectra and blue lines show fitted
        baselines. }
   \label{spec-sed}                
\end{figure*}

Here, we show four examples of the \CII\ spectra and SEDs obtained.
Figure~\ref{spec1} shows the only spectrum with a baseline fit of
order 1.  Figure~\ref{spec2} shows the spectrum from the nucleus,
where the lowest rms and highest flux are reached. The highest rms and
also the lowest flux peak of the four spectra is shown in position N61
(fig.~\ref{spec4}) which belongs to the northern, outer N2 region.
The SEDs show the strongest warm dust component in the nucleus
(Fig.~\ref{sed2}) while the weakest warm component is seen in N61
(Fig.~\ref{sed4}).  Cold dust components are similar in the inter-arm
region (Fig.~\ref{sed1}), the nucleus, and in BCLM302
(Fig.~\ref{sed3}). The outer points also show weaker cold dust
emission than the other regions.  

\section{\CII/TIR vs. CO/TIR}
\label{sec-hailey-tir}

\begin{figure*}
\centering
\includegraphics[angle=0,scale=1.]{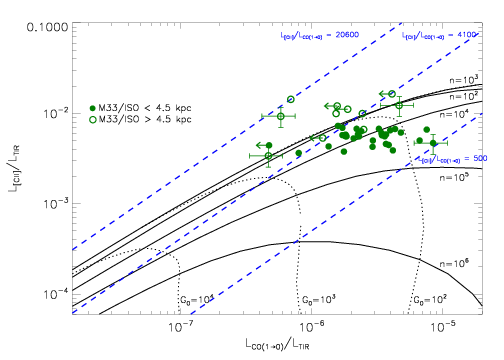}
\caption{\CII\ versus CO, normalized with the total infrared continuum
  (TIR). Big green filled circles show ISO/LWS data of the inner S1,
  N1 regions of M\,33 while open circles show data of the outer S2, N2
  regions.  The lowest \CII/CO ratio observed with ISO/LWS in M\,33 is
  1000 (lower blue dashed line), while the highest ratio is 41200
  (upper blue dashed line).  Black solid and dotted lines indicate
  lines of constant density $n$ and FUV field $G_0$, respectively,
  from the standard K99 PDR model with $A_{\rm {V}}=10$\,mag and solar
  metallicity $Z=1$. }
\label{fig-hailey-tir}
\end{figure*}

For completeness, we show in Figure\,\ref{fig-hailey-tir} the
diagnostic plot of \CII\ vs. CO luminosities, normalized with the TIR
luminosities derived from the two-greybody fits described in
Section\,\ref{FIR-obs}.

\end{appendix}

\end{document}